\documentclass[%
 superscriptaddress,
 reprint,
 amsmath,
 amssymb,
 prl,
 showpacs,
 twocolumn,
 floatfix,
 nofootinbib,
]{revtex4-1}

\usepackage{graphicx}
\usepackage{dcolumn}
\usepackage{bm}
\usepackage{hyperref}
\usepackage{color}
\usepackage{braket}
\usepackage{url}
\usepackage{xcolor}
\usepackage{soul}
\usepackage{lineno}

\begin{document}

\title{Beyond-DFT $\textit{ab initio}$ Calculations for Accurate Prediction of Sub-GeV Dark Matter Experimental Reach}

\author{Elizabeth A. Peterson} \affiliation{Theoretical Division, Los Alamos National Laboratory, Los Alamos, NM 87545, USA}
\author{Samuel L. Watkins} \affiliation{Physics Division, Los Alamos National Laboratory, Los Alamos, NM 87545, USA}
\author{Christopher Lane} \affiliation{Theoretical Division, Los Alamos National Laboratory, Los Alamos, NM 87545, USA}
\author{Jian-Xin Zhu} \affiliation{Theoretical Division, Los Alamos National Laboratory, Los Alamos, NM 87545, USA} \affiliation{Center for Integrated Nanotechnologies, Los Alamos National Laboratory, Los Alamos, NM 87545, USA}

\begin{abstract}
As the search space for light dark matter (DM) has shifted to sub-GeV DM candidate particles, increasing attention has turned to solid state detectors built from quantum materials. While traditional solid state detector targets (e.g. Si or Ge) have been utilized in searches for dark matter (DM) for decades, more complex, anisotropic materials with narrow band gaps are desirable for detecting sub-MeV dark matter through DM-electron scattering and absorption channels. In order to determine if a novel target material can expand the search space for light DM it is necessary to determine the projected reach of a dark matter search conducted with that material in the DM mass – DM-electron scattering cross-section parameter space. The DM-electron scattering rate can be calculated from first-principles with knowledge of the loss function, however the accuracy of these predictions is limited by the first-principles level of theory used to calculate the dielectric function. Here we perform a case study on silicon, a well-studied semiconducting material, to demonstrate that traditional Kohn-Sham density functional theory (DFT) calculations erroneously overestimate projected experimental reach. We show that for silicon this can be remedied by the incorporation of self-energy corrections as implemented in the GW approximation. Moreover, we emphasize the care that must taken in selecting the appropriate level of theory for predicting experimental reach of next-generation complex DM detector materials.
\end{abstract}

\maketitle

\section{\label{sec:intro}Introduction}

As the search for dark matter (DM) particles turns towards efforts to detect light DM in the sub-GeV mass range, solid-state detectors have increasingly become detector targets of interest ~\cite{Essig2016, Hochberg2016A, Hochberg2016B, Hochberg2016C, Hochberg2018, Griffin2018, Geilhufe2020, Griffin2021A, Hochberg2021, Knapen2021A, Coskuner2021, Kahn2022, Huang2023}. Recent experimental and theoretical efforts have focused on absorption and scattering of bosonic and fermionic light DM particles with nucleons, electrons, and bosonic quasiparticles like phonons in detector materials ~\cite{Hochberg2016A, Hochberg2016B, Hochberg2016C, Hochberg2018, Griffin2018, Trickle2020B, Griffin2021A, Hochberg2021, Knapen2021A, Coskuner2021}. As nuclear recoil experiments have struggled to detect GeV-scale DM candidate particles, attention has turned to electron recoil experiments in semiconductor targets, that promise to enable detection of even lighter DM due to their meV-eV electronic excitation energies ~\cite{Essig2016, Hochberg2018, Geilhufe2020, Kahn2022}.

Next generation DM detectors that can probe even lighter DM will require more complex materials such as topological materials, superconductors, etc. which host desirable properties such as narrow band gaps and anisotropies that capture the daily modulation in the DM wind impinging on the Earth ~\cite{Hochberg2016A, Hochberg2016B, Hochberg2016C, Hochberg2018, Geilhufe2020, Rosa2020, Hochberg2021, Coskuner2021}. The novelty and complexity of these materials, however, may limit the experimental ease of rapid characterization and screening for desirable properties. As such, leveraging first-principles methods that handle many-body effects and strong correlations will become increasingly important in predicting projected experimental reach of novel DM detector materials. Moreover, the calculated DM-electron scattering rates may be strongly affected by the first-principles approximations employed with certain, generally low-cost, approximations producing results that may significantly misrepresent the parameter space that is excluded.

In a semiconducting detector, dielectric screening plays a significant role in determining the measured signals from the absorption and scattering of DM particles with electrons ~\cite{Hochberg2021, Knapen2021A, Coskuner2021}. The loss function, i.e. the imaginary part of the inverse dielectric function, can be utilized to predict expected DM-electron scattering rates ~\cite{Nozieres1959, Hochberg2021}, in analogy to electron energy loss spectroscopy (EELS). As such, the complex dielectric function is a central quantity of interest.

From a first-principles perspective, calculation of the dielectric function is a standard procedure ~\cite{Baroni1986, Gajdos2006}. From density functional theory (DFT) calculations, the dielectric function in the random phase approximation (RPA), neglecting local field effects, or exchange-correlation effects, is readily calculable from DFT wavefunctions and eigenvalues. Other methods for calculating the electronic structure or the dielectric function that can capture exchange and (strong) correlation effects include dynamical mean-field theory (DMFT) ~\cite{Metzner1989}, many-body perturbation theory (MBPT) ~\cite{Hedin1965}, density functional perturbation theory (DFPT) ~\cite{Gonze1997}, and time-dependent density functional theory (TD-DFT) ~\cite{Runge1984}. In practice, the methods employed by existing open-source software packages for calculating dielectric functions from first-principles are oriented towards understanding optical properties of materials, in particular absorption of photons in the visible spectrum, and hence restricted to calculations in the long wavelength limit with negligible momentum transfer $\mathbf{q}$. Impinging DM particles may impart both finite momentum $\mathbf{q}$ and finite energy $\omega$ when scattering off an electron, necessitating methods for calculating the dielectric function at finite momentum transfer $\mathbf{q}$.

Response functions like the dielectric function describe the reaction of a material to a perturbation, making them fundamentally excited state properties which require a careful consideration of many-body effects to accurately calculate. In a first-principles framework, in order to accurately capture many-body effects it is necessary to work beyond the standard first-principles DFT formalism. Green's function approaches have been shown to significantly improve upon the accuracy of the computed energy eigenvalues by implementing self-energy corrections ~\cite{Hedin1965, Hybertsen1986, Zhang1989}. The workhorse method of Green's function approximations is the GW approximation ~\cite{Hedin1965, Hybertsen1986} which approximates the self-energy as a product of the single-particle Green's function ($G$) and the screened Coulomb potential ($W$), or $\Sigma=iGW$, neglecting vertex corrections. The GW approximation is typically implemented as a correction scheme to the DFT eigenvalues and wavefunctions to compute quasiparticle-corrected single-particle energy eigenvalues. Beyond improving upon the accuracy of the electronic structure energy eigenvalues, the GW approximation is a convenient method for calculations of the dielectric function. This is because, in GW, calculation of the inverse dielectric function at finite $\mathbf{q}$ is a necessary prerequisite for calculating the screened Coulomb potential ($W$). A number of GW approximation software packages exist ~\cite{Hybertsen1986, Deslippe2012, Marini2009, Sangalli2019, Govoni2015, Gonze2020, Bruneval2006, Shishkin2006} and it has been shown that the GW quasiparticle-correction scheme can immensely improve the accuracy of calculated dielectric functions ~\cite{Hybertsen1986, Zhang1989}. 

There are existing packages, such as \textsc{QEDark}~\cite{Essig2016}, \textsc{DarkELF}~\cite{Knapen2021B}, and \textsc{EXCEED-DM}~\cite{Griffin2021B,Trickle2023}, that enable the calculation of DM-electron scattering rates for various common semiconducting materials used as DM detectors (e.g. Si and Ge). While \textsc{QEDark} does not include in-medium screening effects, \textsc{DarkELF} and \textsc{EXCEED-DM} calculate the scattering based on the dielectric function, thereby including these effects. However, these packages base their dielectric function calculations on DFT outputs, which may not provide the most accurate energy eigenvalues, as noted above. These packages have been frequently used in high-impact results in the field of light DM~\cite{SuperCDMS2018,SuperCDMS2020,SENSEI2018,SENSEI2019,SENSEI2020,DAMIC2019,DAMIC-M2023,CDEX2022,EDELWEISS2020}, suggesting that if there is a significant difference in the DM-electron scattering rate when including, for example, the GW-corrected energy eigenvalues, then these results may be misrepresenting the parameter space that is excluded.

Here we illustrate and emphasize the significance of incorporating many-body effects into calculations of the dielectric function for prediction of projected experimental reach of novel DM detector materials. We present a case study of the well-studied semiconductor silicon (Si), demonstrating that even for a material that does not host substantial relativistic effects or strong correlations, the incorporation of many-body effects is essential for accurate prediction of the projected experimental reach obtained via first-principles calculations. We leave calculation of more complex materials to future work.

\section{\label{sec:theory}Theoretical Framework}

The loss function $L(\mathbf{q},\omega)=\operatorname{Im}\left[-\epsilon(\mathbf{q},\omega)^{-1} \right]$ is the central quantity describing how a dielectric medium responds to an impinging particle. To predict the experimental reach of a detector material, we calculate the electronic structure in order to calculate the complex dielectric function and from there the DM-electron scattering rate and projected reach.

We begin with a DFT calculation using a plane-wave basis to obtain the Kohn-Sham energy eigenvalues and wavefunctions using the Kohn-Sham equations ~\cite{Kohn1965} 

\begin{widetext}
\begin{equation}
H^{KS}\psi_{n\mathbf{k}}^{KS}(\mathbf{r}) = \left[ -\frac{1}{2}\nabla^2 + v_{ext}(\mathbf{r}) + \int \frac{\rho(\mathbf{r')}}{|\mathbf{r}-\mathbf{r'}|}d\mathbf{r'} + v_{XC}[\rho] \right]\psi_{n\mathbf{k}}^{KS}(\mathbf{r}) = \varepsilon_{n\mathbf{k}}^{KS}\psi_{n\mathbf{k}}^{KS}(\mathbf{r}) \;,
\label{eq:ks_dft}
\end{equation}
\end{widetext}
Here energy is measured in atomic units,  $\psi_{n\mathbf{k}}^{KS}(\mathbf{r})$ and $\varepsilon_{n\mathbf{k}}^{KS}$ are Kohn-Sham wavefunctions and eigenvalues parametrized by band index $n$ and crystal momentum $\mathbf{k}$, $\rho(\mathbf{r}) = \sum_{n\mathbf{k}} |\psi_{n\mathbf{k}}(\mathbf{r})|^{2}$ is the charge density, $v_{ext}(\mathbf{r})$ is the external potential from the ionic background, and $v_{XC}[\rho]$ is the exchange-correlation potential encompassing all many-body effects not captured by the single-particle framework.

We employ the GW approximation as a correction scheme to the Kohn-Sham DFT energy eigenvalues by replacing the exchange-correlation potential with the self-energy operator as in Refs. ~\cite{Hedin1965, Hybertsen1986}:
\begin{widetext}
\begin{equation}
\left[ -\frac{1}{2}\nabla^2 + v_{ext}(\mathbf{r}) + \int \frac{\rho(\mathbf{r')}}{|\mathbf{r}-\mathbf{r'}|}d\mathbf{r'} + \Sigma(\varepsilon_{n\mathbf{k}}^{GW}) \right]\psi_{n\mathbf{k}}^{GW}(\mathbf{r}) = \varepsilon_{n\mathbf{k}}^{GW}\psi_{n\mathbf{k}}^{GW}(\mathbf{r})\;,
\label{eq:gw_eq}
\end{equation}
\end{widetext}

Specifically, we perform a single-shot $G_{0}W_{0}$ calculation, using the Kohn-Sham DFT eigenvalues and wavefunctions to produce the self-energy operator and then solve for the GW energy eigenvalues $\varepsilon_{n\mathbf{k}}^{GW}$ as
\begin{widetext}
\begin{equation}
H^{KS}\psi_{n\mathbf{k}}^{KS}(\mathbf{r}) - 
\left[ \Sigma(\varepsilon_{n\mathbf{k}}^{KS}) - v_{XC}[\rho] \right]\psi_{n\mathbf{k}}^{KS}(\mathbf{r}) = \varepsilon_{n\mathbf{k}}^{GW}\psi_{n\mathbf{k}}^{KS}(\mathbf{r}) \;.
\label{eq:g0w0_eq}
\end{equation}
\end{widetext}
With the Kohn-Sham DFT wavefunctions $\psi_{n\mathbf{k}}^{KS}$ and GW energy eigenvalues $\varepsilon_{n\mathbf{k}}^{GW}$ we may construct the complex dielectric function.

Most generally, the dielectric tensor is a linear response function that describes how a displacement field in a dielectric medium is produced by application of a perturbing electric field as

\begin{equation}
    D_{\alpha}(\mathbf{q},\omega) = \sum_{\beta}\epsilon_{\alpha\beta}(\mathbf{q},\omega)E_{\beta}(\mathbf{q},\omega)\;.
\end{equation}

\noindent where $\alpha$ and $\beta$ are Cartesian directions. Several symmetries can be leveraged to simplify the dielectric tensor. In an isotropic material, the longitudinal (diagonal) and transverse (off-diagonal) components of the dielectric tensor can be strictly separated. In a material with cubic symmetry, the diagonal components of the dielectric tensor reduce to a single scalar function, significantly simplifying the problem of calculating the dielectric function.

We follow the standard procedure to calculate the scalar dielectric function in the random phase approximation (RPA) in reciprocal space as:

\begin{equation}
    \epsilon_{\mathbf{GG'}}(\mathbf{q},\omega)=\delta_{\mathbf{GG'}}-v(\mathbf{q+G})P_{\mathbf{GG'}}(\mathbf{q},\omega) \;,
    \label{eq:eps_ggp}
\end{equation}
 where $v$ is the bare Coulomb potential, $P$ is the polarizability, $\mathbf{q}$ is a wave vector confined to the first Brillouin zone (BZ), and $\mathbf{G}$ is a reciprocal lattice vector. For an isotropic material in the RPA, the polarizability $P$ can be expressed as the independent particle polarizability in terms of matrix elements of the density operator, where $\rho_{\mathbf{q}}=\sum_{i} e^{-i \mathbf{q}\cdot\mathbf{x}_{i}}$ is the density operator for a system of electrons at positions ${\mathbf{x}_{i}}$ in the presence of a momentum transfer $\mathbf{q}$ from a fast impinging charged particle ~\cite{Nozieres1959}. The polarizability is then written as

\begin{widetext}
\begin{equation}
    P_{\mathbf{GG'}}(\mathbf{q},\omega)= \operatorname{lim}_{\eta \rightarrow 0} \sum_{nn'\mathbf{k}}\braket{n\mathbf{k}|e^{-i(\mathbf{q}+\mathbf{G})\cdot\mathbf{r}}|n'\mathbf{k+q}}\braket{n'\mathbf{k+q}|e^{i(\mathbf{q}+\mathbf{G'})\cdot\mathbf{r}}|n\mathbf{k}}\frac{f(\varepsilon_{n'\mathbf{k+q}}) - f(\varepsilon_{n\mathbf{k}})}{\varepsilon_{n'\mathbf{k+q}} - \varepsilon_{n\mathbf{k}} - \hbar\omega -i\hbar\eta}\;,
\end{equation}
\end{widetext}
where $n$ is a band index of an electronic state, $\mathbf{k}$ is the crystal momentum of an electronic state, $\mathbf{q}$ is the momentum transfer (here restricted to the first Brillouin zone), $\mathbf{G}$ is a reciprocal lattice vector, $f(\varepsilon)$ is the Fermi-Dirac distribution function, and $\eta$ is an infinitesimal positive quantity.

\begin{figure*}
   \includegraphics[width=0.8\linewidth]{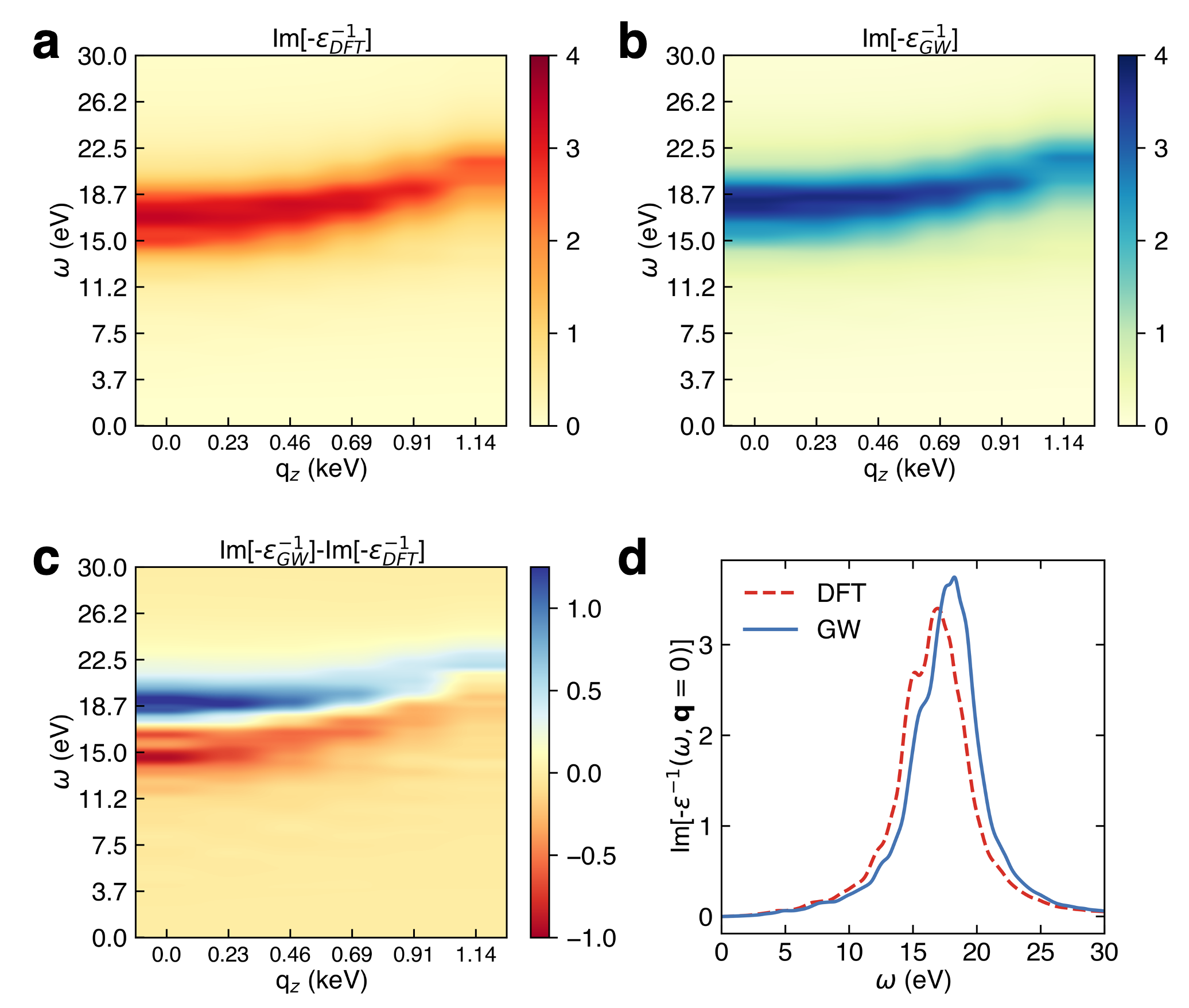}
\caption{\label{fig:imag_dft_gw}The loss function, the negative imaginary part of the inverse dielectric function, for Si calculated with (a) DFT and (b) the GW approximation plotted as a function of energy transfer $\omega$ and momentum transfer $\mathbf{q}$, restricted to the first BZ, in units of energy. The difference between the two loss functions indicates a blue shift in the quasiparticle-corrected GW loss function relative to that calculated using DFT as seen over (c) all the $\mathbf{q}$ points sampled along the $q_{z}$ axis of the first BZ and in (d) the $\mathbf{q}\rightarrow 0$ cut.}
\end{figure*}

\begin{figure*}
   \includegraphics[width=0.8\linewidth]{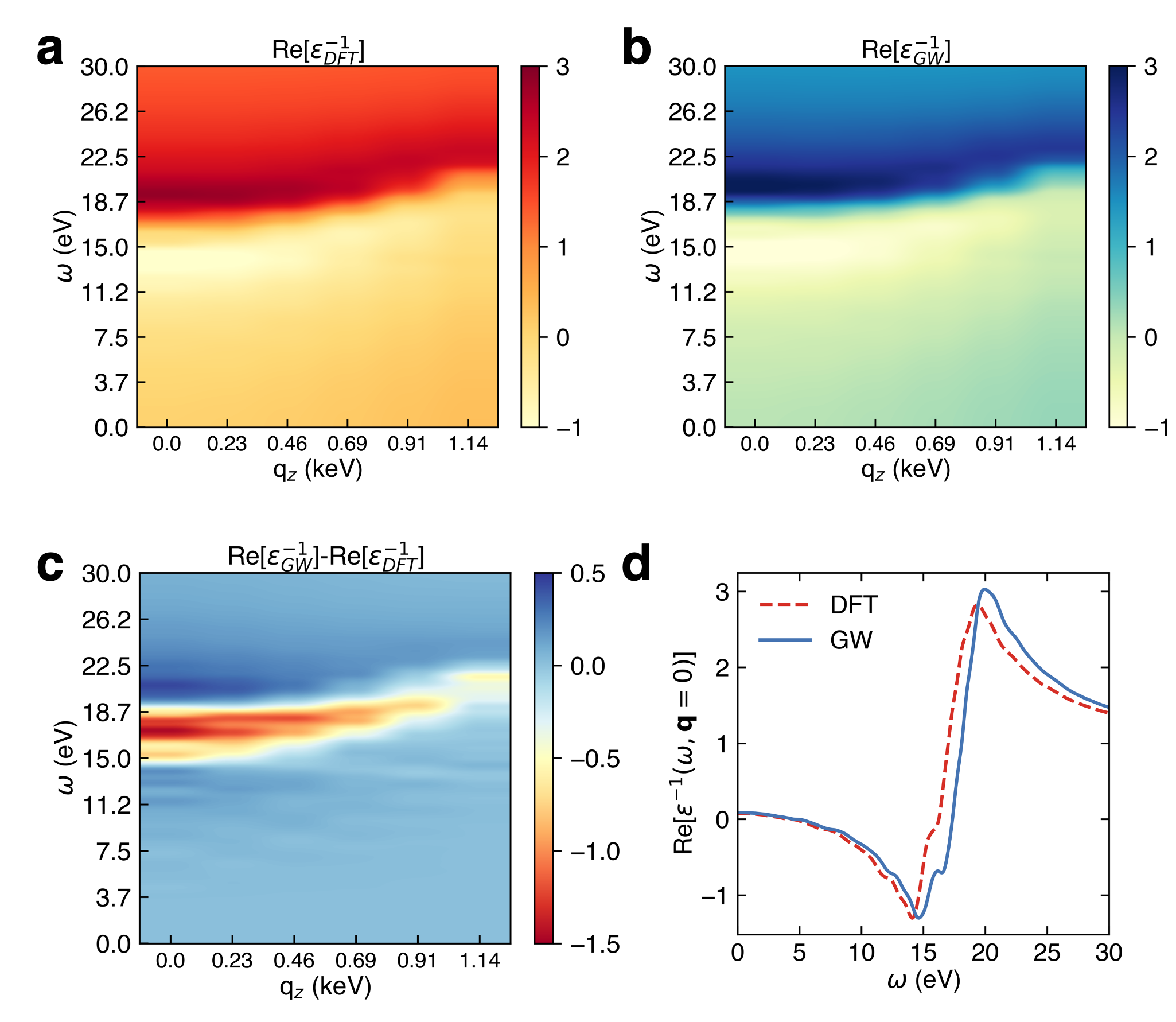}
\caption{\label{fig:real_dft_gw}The real part of the inverse dielectric function for Si calculated with (a) DFT and (b) the GW approximation plotted as a function of energy transfer $\omega$ and momentum transfer $\mathbf{q}$, restricted to the first BZ, in units of energy. The difference between the two inverse dielectric functions indicates a blue shift in the quasiparticle-corrected GW dielectric function relative to that calculated using DFT as seen over (c) all the $\mathbf{q}$ points sampled along the $q_{z}$ axis of the first BZ and in (d) the $\mathbf{q}\rightarrow 0$ cut.}
\end{figure*}

The DM-electron scattering rate can be written in terms of the loss function $L(\mathbf{q},\omega)=\operatorname{Im}\left[-\epsilon(\mathbf{q},\omega)^{-1} \right]$ ~\cite{Hochberg2021, Knapen2021A}, or in some formalisms the dynamic structure factor, a related quantity ~\cite{Griffin2020,Trickle2020A, Sturm1991}. For example, for an impinging DM particle $\chi$ with velocity $\mathbf{v}_{\chi}$ and an arbitrary DM-electron interaction potential $V(\mathbf{q})$, Ref. ~\cite{Hochberg2021} outlines how the transition rate can be expressed as
\begin{equation}
    \Gamma(\mathbf{v}_{\chi})=\int \frac{d^{3}q}{(2\pi)^{3}}|V(\mathbf{q})|^{2} \left[ 2\frac{q^{2}}{e^{2}} \operatorname{Im}\left(-\frac{1}{\epsilon(\mathbf{q},\omega)} \right)\right]\;.
    \label{eq:transition_rate}
\end{equation}

From the first-principles calculated inverse dielectric function, we estimate the expected sensitivity of a detector to DM-electron scattering following the formalism outlined in Refs.~\cite{Trickle2020A, Geilhufe2020, Hochberg2021, Knapen2021A, Kahn2022}. The total rate per target mass at time $t$ can be expressed as
\begin{equation}
    R(t) = \frac{1}{\rho_T}\frac{\rho_\chi}{m_\chi} \int d^3\mathbf{v}_\chi f_\chi(\mathbf{v}_\chi, t) \Gamma(\mathbf{v}_\chi),
    \label{eq:dm_scattering_general}
\end{equation}
where the transition rate $\Gamma(\mathbf{v}_\chi)$ has been defined in Eq.~(\ref{eq:transition_rate}), $f_\chi(\mathbf{v}_\chi, t)$ is the DM velocity distribution at time $t$, $\rho_\chi = 0.3\, \mathrm{GeV}/\mathrm{cm}^3$ is the local DM density, $m_\chi$ is the DM mass, and $\rho_T$ is the mass density of the target material.

In DM-electron scattering, the scattering rate is usually parameterized in the literature with
\begin{align}
    V(\mathbf{q}) &= \frac{g_\chi g_e}{q^2 + m_{V/\phi}^2}, \\
    \bar{\sigma}_e &\equiv \frac{\mu_{\chi e}^2}{\pi} \left|V(q_0)\right|^2,\\
    F_\mathrm{DM}(q) &\equiv \frac{q_0^2 + m_{V/\phi}^2}{q^2 + m_{V/\phi}^2}, 
\end{align}
where the DM is coupled to the Standard Model through some scalar ($\phi$) or vector ($V$) mediator with mass $m_{V/\phi}$ with coupling $g_\chi$ to DM and coupling $g_e$ to electrons. Here, $\bar{\sigma}_e$ is a reference cross section for DM-electron scattering, $\mu_{\chi e}$ is the DM-electron reduced mass \mbox{$\mu_{\chi e}=\frac{m_{e}m_{\chi}}{m_{e}+m_{\chi}}$}, \mbox{$q_0\equiv \alpha m_e$} is a reference momentum, and $F_\mathrm{DM}(q)$ is the DM form factor. With Eqs.~(\ref{eq:transition_rate}) and (\ref{eq:dm_scattering_general}), we arrive at the following scattering rate per target mass at time $t$
\begin{widetext}
\begin{equation}
    R(t) = \frac{\bar{\sigma}_e}{\rho_T} \frac{\rho_\chi}{m_\chi} \frac{\pi}{(2\pi)^4 \alpha \mu_{\chi e}^2} \int d\omega \, d^3\mathbf{q}\, q \, F_\mathrm{DM}^2(q) \operatorname{Im}\left[-\frac{1}{\varepsilon(\mathbf{q}, \omega_\mathbf{q})}\right] \tilde{g}(v_\mathrm{min}, \psi, t)\;,
\end{equation}
\end{widetext}
where we have defined
\begin{equation}
    \tilde{g}(v_\mathrm{min}, \psi, t) = q \int d^3\mathbf{v}_\chi f_\chi(\mathbf{v}_\chi, t) \delta(E_f - E_i)\;,
\end{equation}
similarly to Ref.~\cite{Geilhufe2020}.

To calculate the expected sensitivity of a Si detector using our DFT and GW calculations, we assume a background-free search with a kg-year exposure, and set limits at the 90\% C.L. We assume the Standard Halo Model~\cite{Drukier1986,Evans2018} and average the time-dependent rate over the full exposure (a small effect for an isotropic crystal such as Si).

Two caveats of this formalism are immediately apparent. The first is that it is highly dependent on the accuracy of the first-principles wavefunctions and energy eigenvalues used to compute the polarizability. For highly accurate predictions of scattering rates this requires using first-principles methods beyond traditional density functional theory (DFT). The second is that it is restricted to calculation of scalar dielectric functions, limiting its ability to capture screening effects in highly anisotropic materials. For this work we address the first of these caveats, leaving discussion of anisotropic materials to forthcoming work.

\section{\label{sec:calc_details}Calculation Details}

To calculate the finite $\mathbf{q}$ RPA dielectric function of Si, we first perform density functional theory calculations using Quantum Espresso ~\cite{Gianozzi2009, Gianozzi2017, Gianozzi2020}. We use the BerkeleyGW package ~\cite{Hybertsen1986, Deslippe2012} to calculate the full frequency dielectric function using DFT-level energy eigenvalues and wavefunctions. We next calculate the self-energy operator and quasiparticle corrections in the GW approximation and recalculate the full frequency dielectric function with the quasiparticle-corrected GW energy eigenvalues.

DFT calculations are performed with a plane-wave basis in the generalized-gradient approximation (GGA) as implemented by Perdew, Burke, and Ernzerhof (PBE) ~\cite{Perdew1996} using a 100 Ry energy cut-off on a $10\,\times10\,\times10$ $\mathbf{k}$-grid. A scalar relativistic norm-conserving pseudopotential including 4 valence electrons per Si atom from the Pseudo-Dojo project is used ~\cite{Hamann2013}.

The DFT-level and quasiparticle-corrected GW full frequency dielectric functions are calculated using a 4.0 Ry $\mathbf{G}$-vector cut-off and 100 meV broadening for a frequency range from 0-30 eV on a $10\,\times10\,\times10$ $\mathbf{q}$-grid. The self-energy operator is calculated using 4 filled and 96 empty bands.

\begin{figure}
    \centering
      \includegraphics[width=1.0\linewidth]{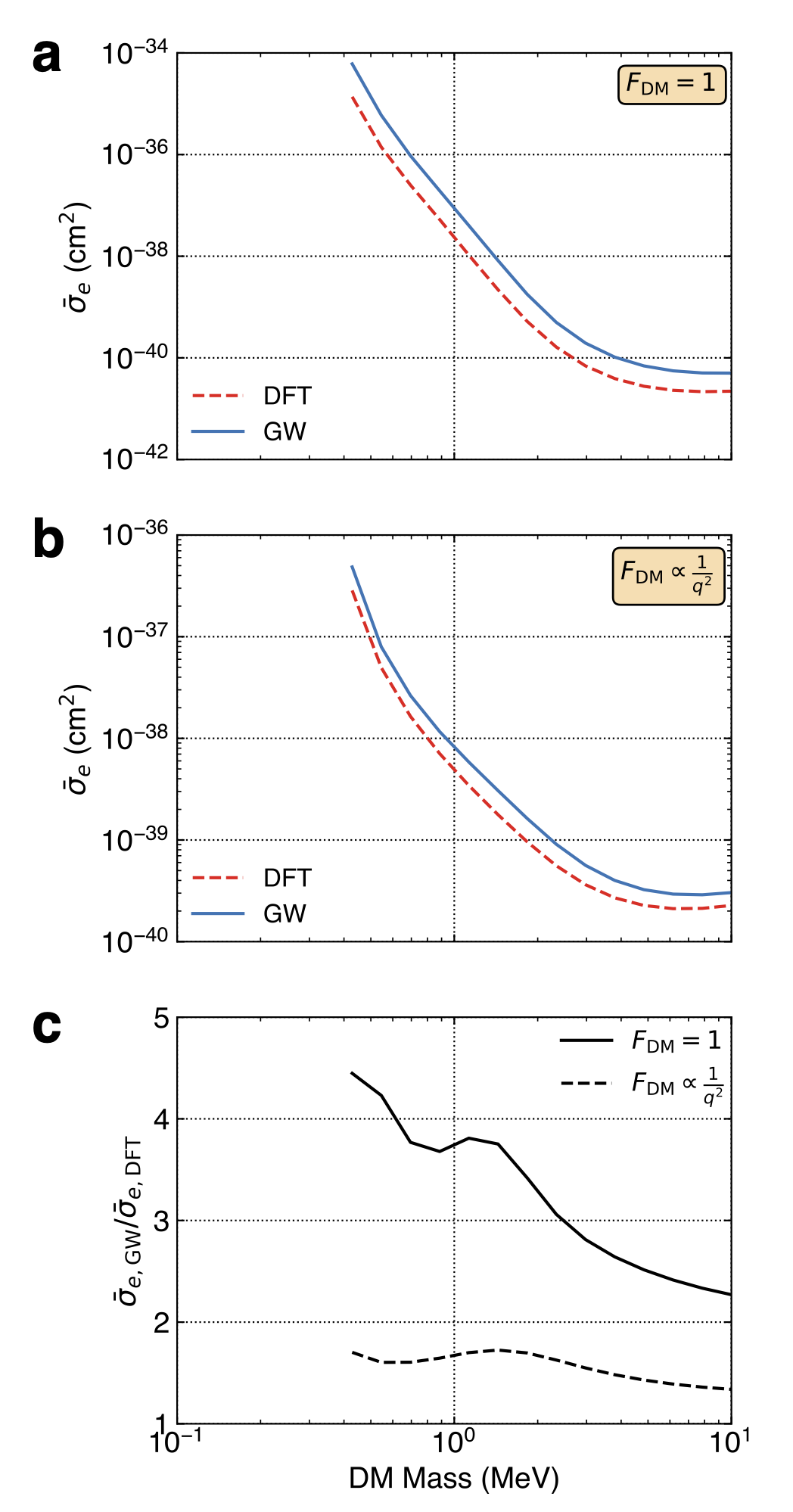}
    \caption{Sensitivity curves (90\% C.L.) for DM-electron scattering with a Si target, assuming a kg-yr exposure and zero background for a (a) heavy mediator and (b) light mediator. In both panels, we compare the sensitivity obtained using the loss function obtained from DFT and that calculated using the GW approximation, where we see a significant difference in the expected sensitivity and (c) relative electron-DM scattering cross-section.}
    \label{fig:si_sens}
\end{figure}

\section{\label{sec:results}Results}

We calculate the RPA dielectric function of Si using DFT and GW eigenvalues for a broad range of relevant energy transfers $\omega$ and momentum transfers $\mathbf{q}$ from impinging particles. We calculate the dielectric function for momentum transfer $\mathbf{q}$ up to a 4.0 Ry (54.4 eV) kinetic energy cut-off for plane waves with momentum $\mathbf{q+G}$ and kinetic energy $\frac{\hbar^{2}|\mathbf{q+G}|^2}{2m_{e}}$ with $\mathbf{q}$ and $\mathbf{G}$ defined as in Eq.\ref{eq:eps_ggp} above. We consider energy transfers $\omega$ up to 30 eV. The static dielectric constant $\epsilon(\mathbf{q\rightarrow 0, \omega=0})$ calculated with GW, 11.46,  agrees much better with the experimental value, 11.7 ~\cite{Dunlap1953}, than that calculated using DFT, 12.86.

In Fig.~\ref{fig:imag_dft_gw} we plot the RPA loss function, restricting ourselves to $\mathbf{q}$ vectors along the high symmetry $q_{z}$ line in the first BZ for visual brevity. The momentum transfer $\mathbf{q}$ is plotted in units of energy. The loss function is qualitatively similar for the DFT (Fig. \ref{fig:imag_dft_gw}a) and GW (Fig. \ref{fig:imag_dft_gw}b) eigenvalues. By plotting the difference between the GW and DFT loss functions (Fig. \ref{fig:imag_dft_gw}c,d), we see that the GW loss function is blue shifted to higher energy relative to the DFT loss function, by $\sim$ 1 eV, a substantial energy difference relative to the DM-electron interaction energy-scales of interest. Additionally, the peak of the loss function calculated from GW is slightly larger than that calculated from DFT due to quasiparticle self-energy effects. The blue shift of the GW loss function is a result of the larger band gap calculated via the GW approximation (0.97 eV) relative to that calculated using DFT (0.58 eV) as a result of quasiparticle self-energy corrections. This band gap increase shifts the imaginary part of the dielectric function to higher energy because the smallest energy difference for a transition between a filled and unfilled band shift to higher energy.

In Fig.~\ref{fig:real_dft_gw} we plot the corresponding real part of the RPA inverse dielectric function. Again, the real part of the RPA inverse dielectric function is qualitatively similar for the DFT (Fig. \ref{fig:real_dft_gw}a) and GW (Fig. \ref{fig:real_dft_gw}b) eigenvalues. The difference between the real parts of the GW and DFT inverse dielectric functions (Fig. \ref{fig:real_dft_gw}c) reveals again that the GW results are blue shifted to higher energy.

In Fig.~\ref{fig:si_sens}, we show the results of the sensitivity calculation for MeV-scale DM masses for the limits of a heavy vector mediator ($m_V\to \infty$) (Fig. ~\ref{fig:si_sens}(a)) and a light vector mediator ($m_V \to 0$) (Fig. ~\ref{fig:si_sens}(b)). The limits calculated with the GW and DFT loss functions are markedly different, emphasizing the importance of accounting for many-body effects in first-principles calculations of the dielectric function for predicting experimental reach. In fact, the bare DFT calculation tends to $\textit{overestimate}$ DM sensitivity. As shown in Fig. ~\ref{fig:si_sens}(c), the electron scattering cross sections corresponding to equivalent DM masses are consistently larger when calculated with GW as compared to DFT. The effect is most significant for heavy mediators, where the electron scattering cross sections can be up to $\sim$4.5 times larger. This corresponds to a substantial overestimation of the parameter space that can be probed by Si when the reach estimation is performed using only DFT. For more complicated, anisotropic crystal structures, we expect this discrepancy to increase and leave those calculations to future work.

\section{\label{sec:summary}Summary}

Here we have used Si as a case study to demonstrate the importance of incorporating beyond-DFT, many-body effects in first-principles calculations of the dielectric function for predicting the expected scattering rates and the experimental reach of detectors utilizing DM-electron scattering for sub-GeV DM detection. We have calculated the RPA dielectric function of Si using DFT eigenvalues and GW-corrected eigenvalues obtained through a single-shot $G_{0}W_{0}$ calculation. By using the loss function to model the response of a Si detector to impinging particles, we have calculated the projected experimental reach of a Si detector. Crucially, we have found that DFT $\textit{overestimates}$ the experimental reach. By using the GW approximation, we have been able to correct for this overestimation to produce a more realistic projected reach. As recent DM search results often rely on DFT to calculate limits on DM-electron scattering, these limits are likely overestimating their exclusion regions. We suggest to revisit these results with energy eigenvalues that have been calculated at a higher level of theory, such as the GW approximation. The novel detector materials that will be at the forefront of the search for sub-GeV DM are expected to host much more complex many-body and correlation effects than Si. We note that the GW approximation is only one of many ways, such as DMFT, to account for many-body effects and that different treatments will be more appropriate for different materials. Ultimately though, our work underscores the limitations of DFT and the importance of utilizing beyond-DFT methods in first-principles calculations of the experimental reach of both current and next-generation sub-GeV DM detectors.

\section{\label{sec:ackn}Acknowledgements}
This work was supported by the U.S. DOE NNSA under Contract No. 89233218CNA000001. E.A.P., C. L. and J.-X.Z. acknowledge support by the LANL LDRD Program through project number 20220135DR. S.L.W. acknowledges support from the LANL Director's Postdoctoral Fellowship award 20230782PRD1. This work was supported in part by the Center for Integrated Nanotechnologies, a DOE Office of Science user facility, in partnership with the LANL Institutional Computing Program for computational resources.  Additional computations were performed at the National Energy Research Scientific Computing Center (NERSC), a U.S. Department of Energy Office of Science User Facility located at Lawrence Berkeley National Laboratory, operated under Contract No. DE-AC02-05CH11231 using NERSC award ERCAP0020494.

\bibliographystyle{aps}
\bibliography{citations}

\end{document}